\documentclass[%
 reprint,
superscriptaddress,
 amsmath,amssymb,
 aps,
]{revtex4-2}

\usepackage{graphicx}
\usepackage{dcolumn}
\usepackage{bm}

\begin{document}

\preprint{APS/123-QED}

\title{Multi-component altermagnet: A general approach to generating multi-component structures with two-dimensional altermagnetism}

\author{Hongjie Peng}
\author{Sike Zeng}
\author{Ji-Hai Liao}
\author{Chang-Chun He}
\email{scuthecc@scut.edu.cn}
\author{Xiao-Bao Yang}
\author{Yu-Jun Zhao}
\collaboration{Department of Physics, South China University of Technology, Guangzhou 510640, China}

\date{\today}

\begin{abstract}
Altermagnetism, as an unconventional antiferromagnetism, exhibits collinear-compensated magnetic order in real space and spin-splitting band structure in reciprocal space. In this work, we propose a general approach to generating multi-component structures with two-dimensional altermagnetism, based on symmetry analysis. Specifically, by analyzing the space group of the crystal structures and their subgroups, we systematically categorize equivalent atomic positions and arrange them into orbits based on symmetry operations. Chemical elements are then allowed to occupy all atomic positions on these orbits, generating candidate structures with specific symmetries. We present a general technique for generating collinear-compensated magnetic order, characterized by the symmetrical interconnection between opposite-spin sublattices, and employ first-principles calculations to determine magnetic ground states of multi-component materials. This approach integrates symmetry analysis with the screening of altermagnetic configurations to evaluate the likelihood of candidates possessing altermagnetism. To verify the methodology, we provide examples of previously unreported 2D altermagnets, such as $\mathrm{Cr}_2 \mathrm{Si}_2 \mathrm{S}_3 \mathrm{Se}_3$, $\mathrm{Fe}_2 \mathrm{P}_2 \mathrm{S}_3 \mathrm{Se}_3$, and $\mathrm{V}_2 \mathrm{O}_2 \mathrm{Br} \mathrm{I}_3$, and evaluate their dynamical stability by calculating the phonon spectrum. The results demonstrate the feasibility of our approach in generating stable multi-component structures with two-dimensional altermagnetism. Our research has significantly enriched the candidate materials for 2D altermagnet and provided a reference for experimental synthesis. 
\end{abstract}

\maketitle


\section{\label{sec.1}INTRODUCTION}

Magnetic moments interact in intricate ways, producing diverse magnetic phases, including conventional ferromagnetism, antiferromagnetism, as well as altermagnetism, which has recently attracted significant attention in condensed matter physics \cite{mazin2024altermagnetism,PhysRevX.12.040501, PhysRevX.12.031042, WOS:001312751500001, PhysRevX.12.040002, PhysRevB.107.L100418, PhysRevB.108.L100402, feng2022anomalous, naka2019spin, bose2022tilted, PhysRevB.99.184432, PhysRevLett.132.036702}. Altermagnetism, as an unconventional antiferromagnetism, exhibits collinear-compensated magnetic order in real space, but it is characterized by broken time-reversal symmetry leading to spin splitting in reciprocal space. Conventional antiferromagnets exhibit spin degeneracy in the band structure because the opposite-spin sublattices are connected through inversional or translational symmetry \cite{PhysRevX.12.031042}. However, in altermagnets, the opposite-spin sublattices are interconnected through crystal symmetry operations (such as rotation, mirror, and glide plane) rather than through translational or inversional symmetries, leading to the disruption of \textit{PT} symmetry \cite{PhysRevX.12.031042}. Recent experimental breakthroughs utilizing angle-resolved photoemission spectroscopy have successfully identified spin-splitting electronic structures in the altermagnet, providing direct evidence for the existence of altermagnetism \cite{PhysRevX.12.011028, WOS:001181488200010}.

Compared to a plenty of three-dimensional altermagnets that have been discovered, two-dimensional intrinsic altermagnets were rarely reported \cite{WOS:001312751500001}. Unlike three-dimensional collinear magnetic materials, the connection between opposite-spin sublattices through $m_z$ (the mirror symmetry through the $xy$ plane) or $C_{2z}$ (twofold rotation around the $z$ axis) protects the spin degeneracy of the nonrelativistic band structure for all k vectors in the whole Brillouin zone in two-dimensional collinear magnetic materials \cite{PhysRevB.110.054406}. This may be one of the reasons why discovering two-dimensional altermagnets are scarce. Therefore, a series of methods for searching for or generating two-dimensional altermagnets have been proposed \cite{WOS:001215200500005, mazin2023inducedmonolayeraltermagnetismmnpsse3, PhysRevB.110.174410}, for instance, high-throughput computational screening \cite{WOS:001215200500005}. Although high-throughput computational screening has been employed, the resultant number of two-dimensional altermagnets remains relatively limited, primarily due to the fact that the majority of two-dimensional magnetic materials possessing collinear magnetic configurations do not fulfill the symmetry prerequisites for altermagnets. Moreover, transition from conventional antiferromagnetism to altermagnetism in two-dimensional magnetic materials can be induced by application of external electric field \cite{mazin2023inducedmonolayeraltermagnetismmnpsse3}. Application of an out-of-plane electric field to $\mathrm{Mn} \mathrm{P} \mathrm{S}_3$ or $\mathrm{Mn} \mathrm{P} \mathrm{Se}_3$ can also break the \textit{PT} symmetry and induce extrinsic two-dimensional altermagnetism. In addition, bilayer stacking is a successful general approach to generating two-dimensional altermagnetism \cite{PhysRevB.110.174410, PhysRevLett.133.166701}. Constructing a bilayer system comprising two single ferromagnetic layers with antiferromagnetic coupling can induce altermagnetism. In this way, the bilayer system satisfies the requirement that the opposite sublattices can be connected by symmetries, but $C_{2z}$ and z-direction translation symmetry between the opposite-spin sublattices are excluded, making the bilayer system easier to possess altermagnetism. 

\begin{figure*}[t]
\centering
\includegraphics[width=\textwidth]{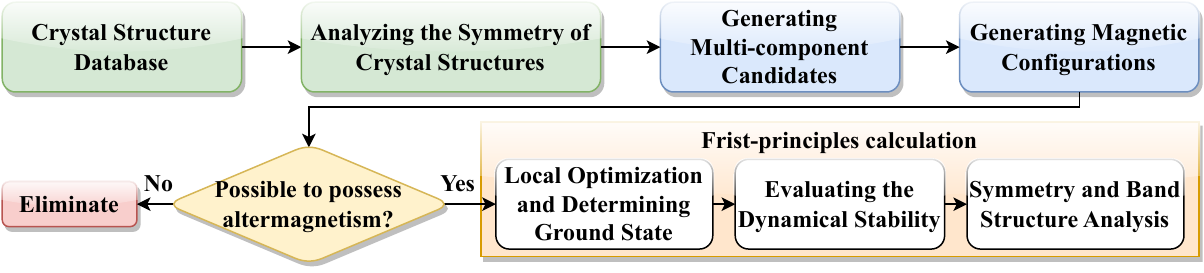}
\caption{Framework for generating stable multi-component altermagnets. We propose a general approach to generating multi-component 2D altermagnetic structures and magnetic configurations based on symmetry analysis of crystal structures. The method combines symmetry analysis with screening of altermagnetic configurations, followed by first-principles calculations to determine magnetic ground states and dynamical stability. Symmetry and band structure analysis confirm altermagnetism in candidate structures. }
\label{Fig.1}
\end{figure*}

The expansion of two-dimensional materials with altermagnetism will greatly enhance our understanding of the fundamental properties of altermagnets and promote their applications in spintronics, for instance, giant and tunneling magnetoresistance (GMR and TMR) effect \cite{PhysRevX.12.011028} can be generated in altermagnets. Therefore, developing an approach to enriching two-dimensional altermagnetic structures is of great significance. Moreover, given the absence of experimentally synthesized two-dimensional altermagnets, it is crucial to establish a reliable method for predicting stable 2D altermagnets.

In this work, we propose a general approach to generating multi-component structures that possibly exhibit two-dimensional altermagnetism, based on group-subgroup relationship and symmetry analysis of crystal structures sourced from Computational 2D Materials Database (C2DB) \cite{Haastrup_2018,Gjerding_2021}. Furthermore, we also propose a general method for generating collinear magnetic configurations characterized by the symmetric interconnection of opposite-spin sublattices. This approach merges symmetry analysis with the screening of altermagnetic configurations to ascertain the potential of candidate structures to possess altermagnetism. Employing first-principles calculations, we conduct local optimization and determine magnetic ground state of these candidates. Following this, we identify structures with dynamical stability by computing of their phonon spectrum. Finally, we conduct a thorough analysis of symmetry and band structure to demonstrate that the candidate structures possess altermagnetism. In conclusion, our work presents a general methodology for generating stable multi-component structures exhibiting two-dimensional altermagnetism, as shown in Fig. \ref{Fig.1}. This contributes to the diversification of 2D altermagnets and offers a reference for the experimental synthesis. 

In our previous study, it has been demonstrated that structures with small number of Wyckoff positions exhibit a great likelihood of being ground state structures of multi-component materials \cite{PhysRevMaterials.6.L050801}. This method  is capable of generating candidate structures with specific symmetries and small number of Wyckoff positions. Therefore, this approach to generating symmetric structures is also beneficial for identifying stable configurations of multi-component materials. 

The article is organized as follows: in Sec. \ref{sec.2}, we present a general approach to generating multi-component structures with altermagnetism. In Sec. \ref{sec.3}, we introduce a method for generating collinear magnetic configurations with features of the connection between opposing sublattices through symmetries. In Sec. \ref{sec.4}, we present examples of previously unreported 2D multi-component altermagnets to validate our approach and analyze their dynamical stability. Finally, we summarize our findings for 2D altermagnetism.

\section{\label{sec.2}AN APPROACH TO GENERATING MULTI-COMPONENT ALTERMAGNET}

\begin{figure*}[t]
\centering
\includegraphics[width=\textwidth]{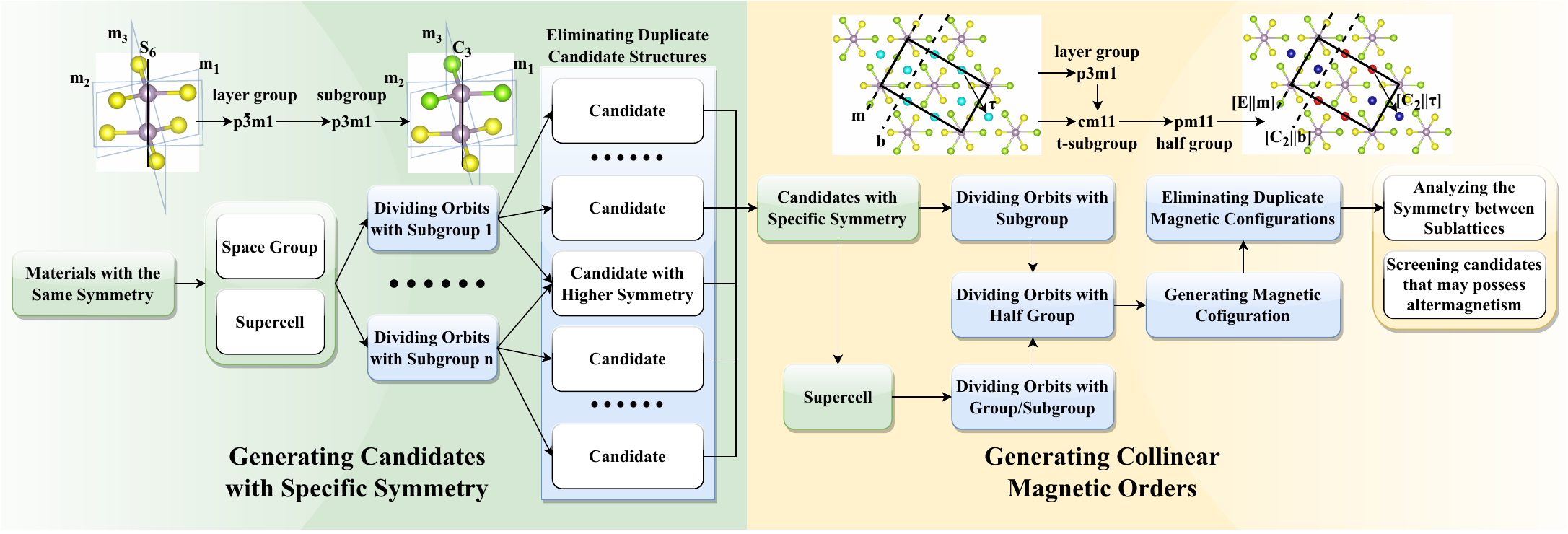}
\caption{An approach to generating multi-component candidates that exhibit particular symmetry requirements of altermagnetism and collinear magnetic orders with features of the connection between opposing sublattices through symmetries. The figure consists of two parts: the left side depicts the approach to generating candidate structures, introduced by Sec. \ref{sec.2}, while the right side depicts the approach to generating collinear magnetic configurations, introduced by Sec. \ref{sec.3}.}
\label{Fig.2}
\end{figure*}

We propose a method for generating multi-component structures possessing specific symmetry, as shown in Fig. \ref{Fig.2}. To construct multi-component altermagnets, which require particular symmetry, it is essential to ensure the candidate structures exhibit specific symmetry. Altermagnetic materials demand that the opposite sublattices occupied by opposite spins are connected by symmetry operations , such as rotational or mirror operations\cite{PhysRevX.12.031042}, which are part of symmetry operations possessed by the crystal structure. Hence, candidates possessing the required symmetry are indispensable for exhibiting altermagnetism. Utilizing the method introduced in this section, we generate a series of candidates fulfilling these symmetry conditions. Initially, we analyze the space (layer) group of a materials class sharing the same symmetry and group-subgroup relationships. Based on primitive cell or supercell, we assign the equivalent atomic positions into different orbits according to the subgroup. By populating these orbits with elements, we generate a series of candidates with specific symmetries. 

The symmetry of crystal structures aligns with a specific space (layer) group, denoted as $G$. Based on primitive cell or supercell, we analyze all subgroups of the  group $G$ and organize them into subgroup chains. Let $X$ be a non-empty set that includes all equivalent atomic positions in the structure, $X=\{x_1,x_2,x_3,\ldots,x_i,\ldots \}$. Let $G$ be the transformation group on $X$. If for all $x_i,x_j \in X$, there exists $g \in G$ such that $g(x_i )=x_j$, meaning there is a crystallographic symmetry operation in $G$ that relates the positions $x_i$ and $x_j$, then $x_i$ and $x_j$ are regarded to be equivalent. The set of all positions in X that are equivalent to the positions $x_i$, which is the collection of all positions that can be mapped to $x_i$ through the symmetry operations of the group $G$, is referred to as the $G$ orbit of $x_i$. When selecting a subgroup $H_k$ from the subgroup chain, divide each class of equivalent atomic positions with subgroup $H_k$ into different orbits. The $H_k$ orbit of general position $x$ is $\{x,h_1 x,h_2 x,\ldots,h_n x\}=\{hx \mid h \in H_k\}$. This process involves categorizing the equivalent atomic positions based on the symmetry operations represented by the subgroup $H_k$. The orbits are essentially the sets of points that can be mapped onto each other through the symmetry operations of the subgroup $H_k$. After obtaining the orbits, we populate these orbits with elements and generate a series of candidates $\{S_1,S_2,\ldots,S_j,\ldots \}$ that satisfy the symmetry requirements of $H_k$. 

We provide a simple proof to demonstrate that the candidate structures $\{S_1,S_2,\ldots,S_j,…\}$ satisfy the symmetry requirements of group $H_k$. Let the position of the $i\textendash th$ atom in the $m\textendash th$ component of the candidate structure $S_j$ be denoted as $\vec{r}_i^{(m)}$. Define the set $R_m=\{ \vec{r}_1^{(m)},\vec{r}_2^{(m)},\ldots,\vec{r}_i^{(m)},\ldots,\vec{r}_n^{(m)} \}$, where $n$ is the number of atoms contained in the m-th component. According to the definition of the subgroup $H_k$ orbit, for all $\vec{r}_i^{(m)} \in R_m$ and $h_\alpha \in H_k$, there exists $\vec{r}_j^{(m)} \in R_m$ such that $\vec{r}_i^{(m)}=h_\alpha \vec{r}_j^{(m)}$, which means $H_k$ keeps $R_m$ invariant. Each component has $R_m=h_\alpha R_m$, where $h_\alpha \in H_k$, which means the subgroup $H_k$ of group $G$ keeps the candidate structure $S_j$ unchanged. When identical components occupy distinct subgroup orbits, they may enable the candidate structure $S_j$ to meet higher symmetry requirements, without impairing the ability of the subgroup $H_k$ of group $G$ to keep the candidate structure $S_j$ unchanged. For instance, identical components occupying different orbits $\{hx \mid h \in H_k\}$ and orbit transformed by $g_\alpha$ as $g_\alpha \{hx \mid h \in H_k\}$ (where $g_\alpha$ belongs to the set $G-H_k$, meaning $g_\alpha$ is an element of $G$ but not of $H_k$), result in additional symmetry operation $g_\alpha$ keeps the candidate structure $S_j$unchanged. Therefore, the symmetry of the candidate structures $\{S_1,S_2,\ldots,S_j,\ldots \}$ correspond to at least the subgroup $H_k$. 

Iterating through all possible ways of components occupying orbits will lead to equivalent candidate structures. Equivalent candidate structures result in redundant analysis and first-principles calculations, so it is necessary to eliminate equivalent structures. Both structures generated by different subgroups and those by the same subgroup may exhibit repetition. If two candidate structures can be connected by a symmetry $g_\alpha$ (where $g_\alpha \in G$), then they are equivalent. Let $S$ be a non-empty set that includes all candidate structures, $S=\{S_1,S_2,…,S_j,\ldots \}$, where the elements $S_1,S_2,S_3,\ldots$ are the candidate structures. Group $G$ is the transformation group on $S$. Since the subgroup $H_k$ of group $G$ keeps the candidate structure $S_j$ unchanged, $H_k=\{h \mid h \in G \text{ and } hS_j=S_j \}$. Then $H_k$ is the isotropy subgroup of $G$ for the candidate structure $S_j$. Therefore, the equivalent structures on the $G$ orbit of the candidate structure $S_j$, $\{S_j,g_1 S_j,g_2 S_j,\ldots,g_n S_j \}$, correspond one-to-one with the left cosets of $H_k$. In this way, we identify and eliminate a part of equivalent candidate structures. 

\section{\label{sec.3}AN APPOARCH TO GENERATING COLLINEAR MAGNETIC CONFIGURATIONS}

After obtaining the atomic arrangement of the magnetic candidate structures, it is necessary to determine the magnetic ground states. To do this, all possible magnetic configurations of the candidate structures must be generated, and then first-principles calculations are used to determine the energy and identify the magnetic ground state. Antiferromagnetic configurations require that the opposite-spin sublattices can be connected by symmetry operations. Therefore, we generate a series of collinear magnetic configurations whose sublattices can be connected by symmetry operations. Based on the symmetry operations that connect opposite sublattices of the altermagnetic configurations in real space, we determined whether the magnetic configurations are possible to possess altermagnetism. If there are configurations with altermagnetism among the collinear magnetic configurations, then it is considered that the candidates are possible to possess altermagnetism. We retain those candidates which have altermagnetic configurations. Then, we use first-principles calculations to determine the energy and find the magnetic ground states of different candidates. In the end, we identify candidates whose ground state is altermagnetic. 

We propose a general method for generating collinear magnetic configurations with features of the connection between opposing sublattices through symmetry operations, and combine it with a method for screening altermagnetic configurations to address whether the candidates possess altermagnetism. The symmetry of the structure corresponds to the space (layer) group $G$, and the set of generators of group $G$ is $\langle f_1,f_2,\ldots,f_m \rangle$. Each individual generator $f_i$ in $\langle f_1,f_2,\ldots,f_m \rangle$ can generate a cyclic group of order $n_i$, and $f_i$ cannot be expressed by $\{f_j \mid j \neq i\}$. The set of generators $\langle f_1,f_2,\ldots,f_{i-1},f_{i+1},\ldots,f_m \rangle$ can produce a group $H$. Since $f_i$ cannot be expressed by $\{f_j \mid j \neq i\}$, $H$ does not contain $f_i,f_i^2,\ldots,f_i^{n_i-1}$, so it is a subgroup of $G$. According to the coset theorem, $f_i H,f_i^2 H,\ldots,f_i^{n_i-1} H$ are obviously disjoint from $H$ and have no common elements. Therefore, the group $G$ can be decomposed into a coset string of subgroup $H$, $G=\{f_i,f_i^2,\ldots,f_i^{n_i}=E\} \bigotimes_\text{S} H$. The notation $\bigotimes_\text{S}$ denotes the semidirect product. Let $x$ be a position occupied by a magnetic atom, $H_x=\{hx \mid h \in H\}$ is the $H$ orbit of $x$, and $G_x=\{gx \mid g \in G\}$ is the $G$ orbit of $x$. Clearly, $H_x \subseteq G_x$. Because the coset strings $f_i H,f_i^2 H,…,f_i^{n_i} H=H$ have no common elements and the same number of elements, the sets $\{gx \mid g \in f_i H\}, \{gx \mid g \in f_i^2 H\}, \ldots, \{gx \mid g \in f_i^{n_i} H\}$ also have no common elements and the same number of elements when $x$ is a general position. Since $G=f_i H \cup f_i^2 H \cup \ldots \cup f_i^{n_i} H$, it follows that 
\begin{equation*}
\begin{split}
&G_x\!=\! \{gx \! \mid \! g \! \in \! G\} \!=\! \{gx \! \mid \! g \! \in \! (f_i H \! \cup \! f_i^2 H \! \cup \! \ldots \! \cup \! f_i^{n_i} H) \} \\
&\!=\! \{gx \! \mid \! g \! \in \! f_i H\} \! \cup \! \{gx \! \mid \! g \! \in \! f_i^2 H\} \! \cup \! \ldots \! \cup \! \{gx \! \mid \! g \! \in \! f_i^{n_i} H\}
\end{split}
\end{equation*}
\begin{figure*}[t]
\centering
\includegraphics[width=\textwidth]{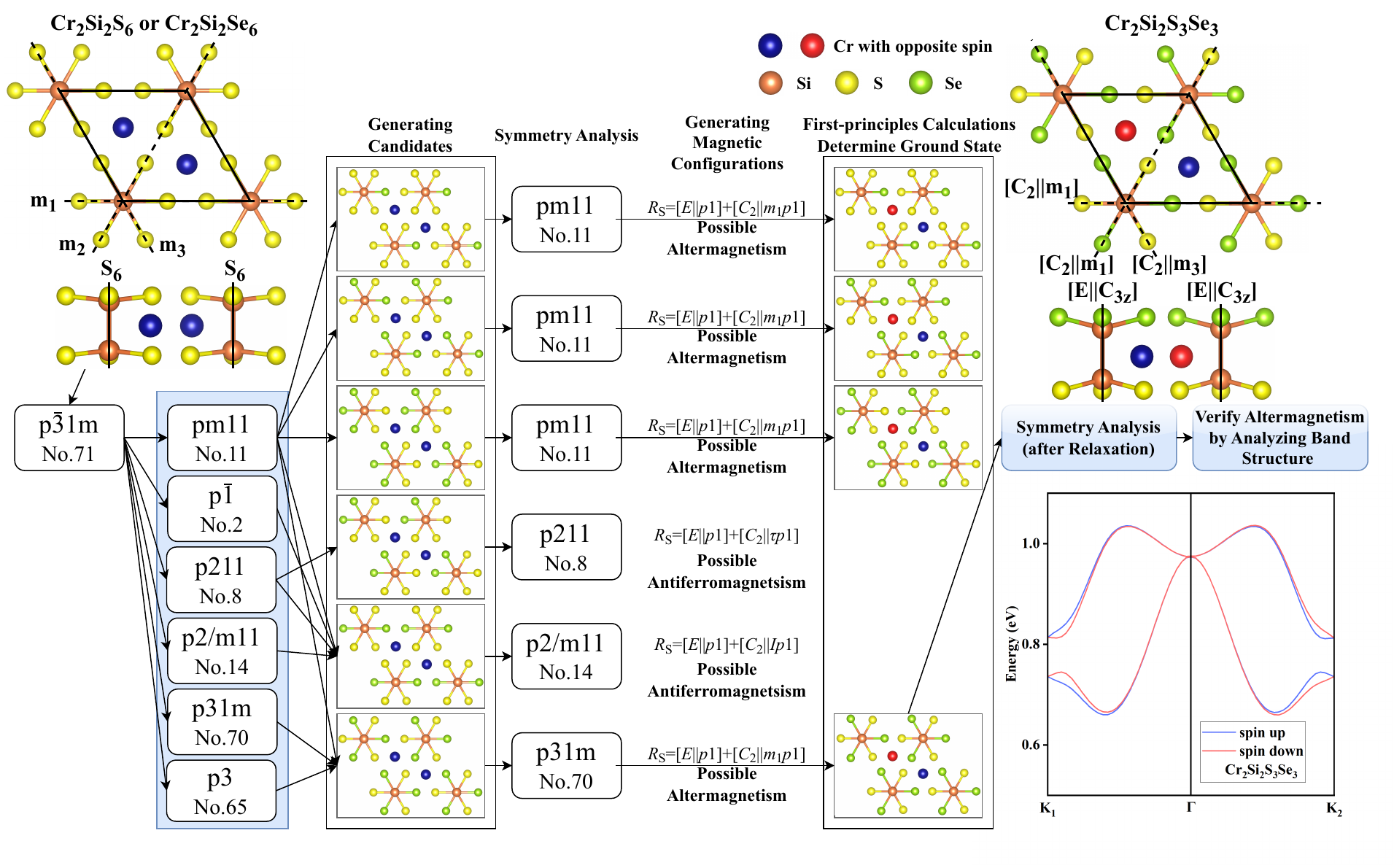}
\caption{An example of searching for altermagnetic structures in the multi-component structures compound $\mathrm{Cr}_2 \mathrm{Si}_2 \mathrm{S}_{6x} \mathrm{Se}_{6(1-x)}$. Blue and red balls represent Cr atoms, with different colors of atoms representing opposite spin sublattices, respectively. Yellow balls represent S atoms, green balls represent Se atoms, or vice versa. Orange balls represent Si atoms. }
\label{Fig.3}
\end{figure*}

For generating collinear magnetic configurations, we focus on the case where $n_i$ is even. When $n_i=2$, the group $G$ is decomposed into coset strings $H,f_i H$. When $n_i$ is even and $n_i>2$, the generators $\langle f_i^2,f_1,f_2,…,f_m \rangle$ (excluding $f_i$) can produce a subgroup $H$. Clearly, the subgroup $H$ does not contain the group element $f_i$. Following this, group $G$ is decomposed into coset strings $H,f_i H$. Therefore, it follows that $G_x=\{gx \mid g \in G\}=\{hx \mid h \in H\} \cup \{gx \mid g \in f_i H\}=\{hx \mid h \in H\} \cup f_i \{hx \mid h \in H\}=H_x \cup f_i H_x$, when $x$ is a general position. When $n_i$ is even, $f_i$ decomposes the $G$ orbit $G_x$ of a general position $x$ into two equal parts with a same number of elements $H_x$ and $f_i H_x$. The subgroup $H$ is referred to as a half group of the group $G$. $H_x$ and $f_i H_x$ are referred to as half orbits. When magnetic atoms occupy $l$ non-equivalent positions, the group $G$ divides the positions of these magnetic atoms into $l$ distinct orbits, each of which corresponds to $l$ pairs of half orbits. From each pair of half orbits, one is chosen to form the set $O_{\mid \uparrow \rangle}$ , where each atomic position $x$ is occupied by an up spin; union of the remaining coset orbits form the set $O_{\mid \downarrow \rangle}$ , where each atomic position $x$ is occupied by a down spin. There are $2^l$ ways to do this. $O_{\mid \uparrow \rangle}$  and $O_{\mid \downarrow \rangle}$  describe the opposite sublattices and magnetic configurations of the structure. This method generates magnetic configurations $\{M_1,M_2,\ldots,M_j,\ldots \}$. 

The union of coset orbits occupied by up spins
\begin{equation*}
O_{\mid \uparrow \rangle}\!=\! \{hx_1 \! \mid \! h \! \in \! H\} \! \cup \! \{hx_2 \! \mid \! h \! \in \! H\} \! \cup \! \ldots \! \cup \! \{hx_l \! \mid \! h \! \in \! H\}
\end{equation*}

The union of coset orbits occupied by down spins
\begin{equation*}
O_{\mid \downarrow \rangle}\!=\! \{gx_1 \! \mid \! g \! \in \! f_iH\} \! \cup \! \{gx_2 \! \mid \! g \! \in \! f_i H\} \! \cup \! \ldots \! \cup \! \{gx_l \! \mid \! g \! \in \! f_i H\}
\end{equation*}
Therefore, 
\begin{equation*}
\begin{split}
&f_i O_{\mid \uparrow \rangle} \\
&\!=\! f_i (\{hx_1 \! \mid \! h \! \in \! H\} \! \cup \! \{hx_2 \! \mid \! h \! \in \! H\} \! \cup \! \ldots \! \cup \{hx_l \! \mid \! h \! \in \! H\})\\
&\!=\! (f_i \! \{hx_1 \! \mid \! h \! \in \! H\}) \! \cup \! (f_i \! \{hx_2 \! \mid \! h \! \in \! H\}) \! \cup \! \ldots \! \cup \! (f_i \! \{hx_l \! \mid \! h \! \in \! H\})\\
&\!=\! \{gx_1 \! \mid \! g \! \in \! f_i H\} \! \cup \! \{gx_2 \! \mid \! g \! \in \! f_i H\} \! \cup \! \ldots \! \cup \! \{gx_l \! \mid \! g \! \in \! f_i H\} \\
&\!=\! O_{\mid \downarrow \rangle}
\end{split}
\end{equation*}

This indicates that $O_{\mid \uparrow \rangle}$  and $O_{\mid \downarrow \rangle}$ can be connected through the symmetry operation $f_i$. When magnetic configuration $M_j^{\prime}$ can be obtained from $M_j$ by spin reversal, these two magnetic configurations are equivalent. Thus, out of the total of $2^l$ ways to choose, there are at most $2^{l-1}$ inequivalent choices. In addition to the symmetry operation $f_i$, there may be other symmetry operations $g_\alpha$ (where $g_\alpha \in f_i H$) such that $O_{\mid \uparrow \rangle}=g_\alpha O_{\mid \downarrow \rangle}$  and $O_{\mid \uparrow \rangle}=g_\alpha O_{\mid \downarrow \rangle}$. In this way, we ensure that there is a symmetry operation connecting the opposite sublattices $O_{\mid \uparrow \rangle}$  and $O_{\mid \downarrow \rangle}$. Therefore, the magnetic configurations described by $O_{\mid \uparrow \rangle}$  and $O_{\mid \downarrow \rangle}$ may have collinear antiferromagnetism or altermagnetism. 

To further determine whether the magnetic configurations described by $O_{\mid \uparrow \rangle}$  and $O_{\mid \downarrow \rangle}$  possess altermagnetism, it is necessary to analyze the symmetry between the sublattices. In three-dimensional materials, when there are symmetry operations between the opposite sublattices whose point group parts are included $\{E,\bar{E} \}$, or in two-dimensional materials, when these elements are included in $\{E,\bar{E},C_{2z},m_z \}$ it results in $\varepsilon (\vec{s},\vec{k})=\varepsilon(-\vec{s},\vec{k})$, leading to spin degeneracy [1]. Consequently, if opposite sublattices in three-dimensional materials possess symmetry operations that encompass $\{E,\tau \} \bigotimes_\text{S} \{E,\bar{E} \}$ or, in two-dimensional materials,  $\{E,\tau \} \bigotimes_\text{S} \{E,\bar{E},C_{2z},m_z \}$, where $\tau$ represents any translation operation, this leads to spin degeneracy. Therefore, the opposite sublattices of altermagnetic orders cannot be linked by the aforementioned symmetry operations. 

For example, when $f_1=C_2$, where $C_2$ is a two-fold rotation operation around an axis in the $xy$ plane, the group $G$ can be decomposed into $G=H \cup C_2 H$, and its orbits can be decomposed into $\{hx \mid h \in H\}$ and $\{gx \mid g \in C_2 H\}$. When up spins occupy the orbit $\{hx \mid h \in H\}$ and down spins occupy $\{gx \mid g \in C_2 H\}$, or vice versa, there is always $O_{\mid \downarrow \rangle}=C_2 O_{\mid \uparrow \rangle}$ . Additionally, $O_{\mid \uparrow \rangle}$  may not be transformed into $O_{\mid \downarrow \rangle}$  through any symmetry operation belonging to the set $\{E,\tau \} \bigotimes_\text{S} \{E,\bar{E} \}$ (and for two-dimensional magnetic materials, this also encompasses the set $\{E,\tau \} \bigotimes_\text{S} \{E,\bar{E},C_{2z},m_z \}$). Therefore, collinear magnetic configuration may possess altermagnetism. If $O_{\mid \uparrow \rangle}$ is able to obtain $O_{\mid \downarrow \rangle}$  through such symmetry operations, the magnetic configuration possesses antiferromagnetism. 

This general approach can generate collinear magnetic configurations, including those with antiferromagnetism and altermagnetism. After obtaining the possible collinear magnetic configurations, we consider the symmetry between the opposite sublattices to determine whether each magnetic configuration possesses altermagnetism. We retain multi-component candidate structures obtained in Sec. \ref{sec.2} that may have altermagnetic configurations among all possible collinear magnetic configurations. Then for each candidate, we use first-principles calculations to determine the energy of different magnetic configurations, and identify its magnetic ground state. By combining symmetry analysis and first-principles calculations, we have discovered several materials with two-dimensional altermagnetism, as presented in Sec. \ref{sec.4}. 

\section{\label{sec.4}PRACTICAL EXAMPLES FOR TWO-DIMENTIONAL MULTI-COMPONENT ALTERMAGNETS}

\begin{figure*}[t]
\centering
\includegraphics[width=\textwidth]{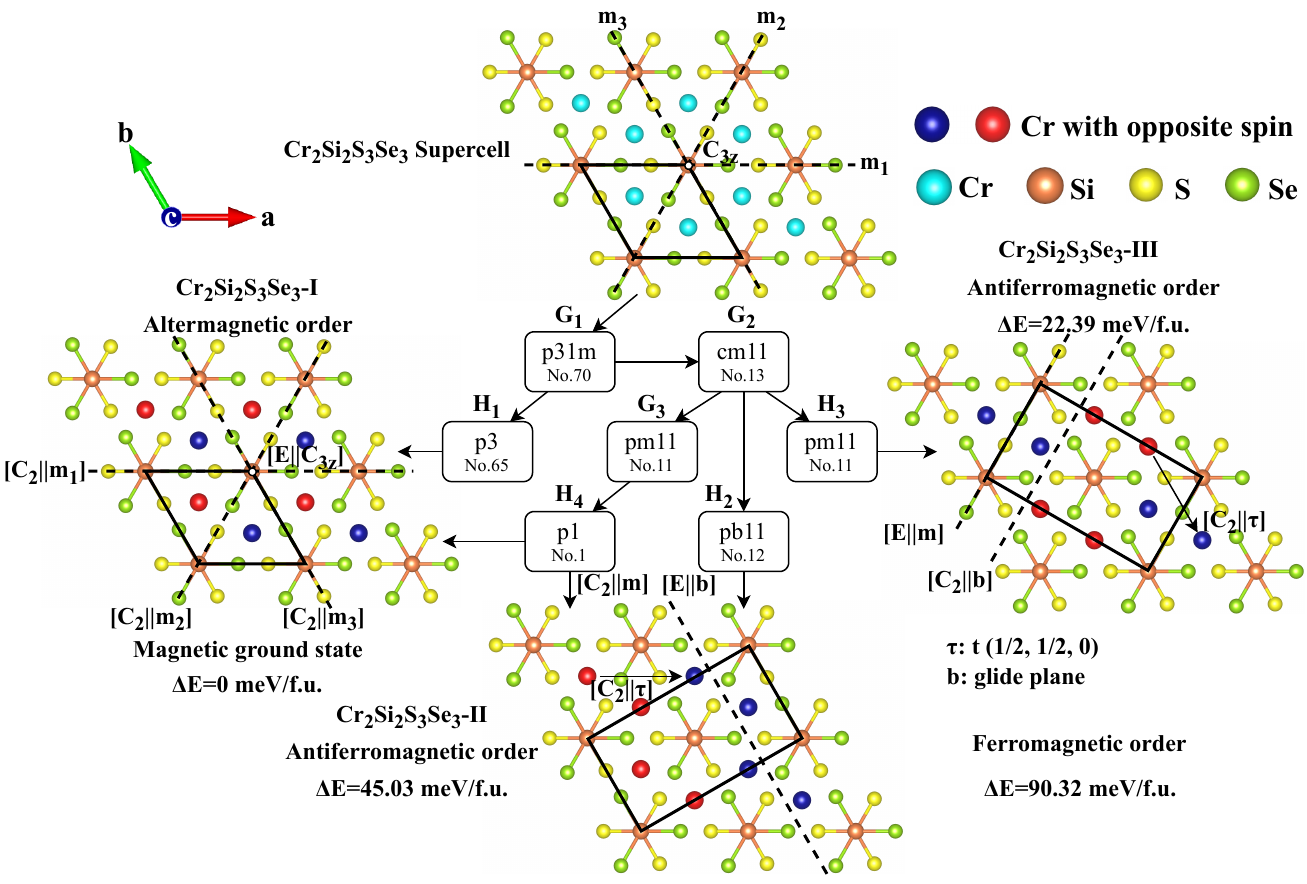}
\caption{An example of generating opposite-spin sublattices in $\mathrm{Cr}_2 \mathrm{Si}_2 \mathrm{S}_3 \mathrm{Se}_3$, which are connected by symmetry operations, is demonstrated based on the methods introduced in Sec. \ref{sec.3}. Cyan, blue, and red balls represent Cr atoms, with red and blue balls representing opposite spin sublattices, respectively. Yellow balls represent S atoms, green balls represent Se atoms, and light purple balls represent P atoms. $\Delta E$ denotes the excess energy per formula unit compared to the magnetic ground state. The notation /f.u. represents averaged to each formula unit.}
\label{Fig.4}
\end{figure*}

In the two-dimensional material database (such as C2DB), classifying materials by layer groups can yield a set of materials with the same symmetry. In such materials, a batch of magnetic materials with similar chemical formulas and structures can be screened out, such as $\mathrm{A}_2 \mathrm{B}_2 \mathrm{C}_6$($\mathrm{Mn}_2 \mathrm{P}_2 \mathrm{S}_6$, $\mathrm{Mn}_2 \mathrm{P}_2 \mathrm{Se}_6$, $\mathrm{Cr}_2 \mathrm{Si}_2 \mathrm{S}_6$, $\mathrm{Li}_2 \mathrm{Mn}_2 \mathrm{Cl}_6$). Materials such as $\mathrm{Cr}_2 \mathrm{Si}_2 \mathrm{S}_6$ and $\mathrm{Cr}_2 \mathrm{Si}_2 \mathrm{Se}_6$ possess similar symmetries, corresponding to the space group $p\bar{3}1m$ (layer group No.71). In these materials, sulfur (S) and selenium (Se) atoms are in equivalent positions. Taking $\mathrm{Cr}_2 \mathrm{Si}_2 \mathrm{S}_{6x} \mathrm{Se}_{6(1-x)}$ as an example, we generate multi-component structures with altermagnetism.  These structures are generated based on the primitive cell, and to produce magnetic configurations, we utilize a 4-times larger cell. Materials $\mathrm{Cr}_2 \mathrm{Si}_2 \mathrm{S}_6$ and $\mathrm{Cr}_2 \mathrm{Si}_2 \mathrm{Se}_6$ with symmetry corresponding to $p\bar{3}1m$ (layer group No.71). We analyze all the subgroups of the layer group $p\bar{3}1m$. The layer group $p\bar{3}1m$ has non-trivial subgroups $p\bar{1}$, $pm11$, $cm11$, $p211$, $c211$, $c2⁄m11$, $p3$,  $p\bar{3}$, $p312$, $p31m$. Corresponding to different subgroups $H_k$, the equivalent positions in the primitive cell are divided into different subgroup $H_k$ orbits. After obtaining the orbits, we populate these orbits with elements and generate a series of candidates with specific symmetries $H_k$. During this process, equivalent structures may be generated, which need to be deduplicated. After eliminating equivalent structures, there are only 12 candidates, some of which are shown in Fig. \ref{Fig.3}. If supercell is considered, the candidates obtained may be more. Based on the methods introduced in Sec. \ref{sec.3}, we generate collinear magnetic configurations for $\mathrm{Cr}_2 \mathrm{Si}_2 \mathrm{S}_{6x} \mathrm{Se}_{6(1-x)}$. By analyzing the symmetry operations that connect the opposite sublattices, we determine whether these structures are possible to possess altermagnetism. Then, we retain candidate structures that may have altermagnetism, narrowing down our selection to 7 candidates. After that, we use first-principles calculations to determine the magnetic ground state of these candidates, thereby judging whether they indeed possess altermagnetism. The ionic relaxation step in DFT calculations may alter the symmetry of the structures, hence a new symmetry analysis is required after the ionic relaxation. 

\begin{figure*}[t]
\centering
\includegraphics[width=\textwidth]{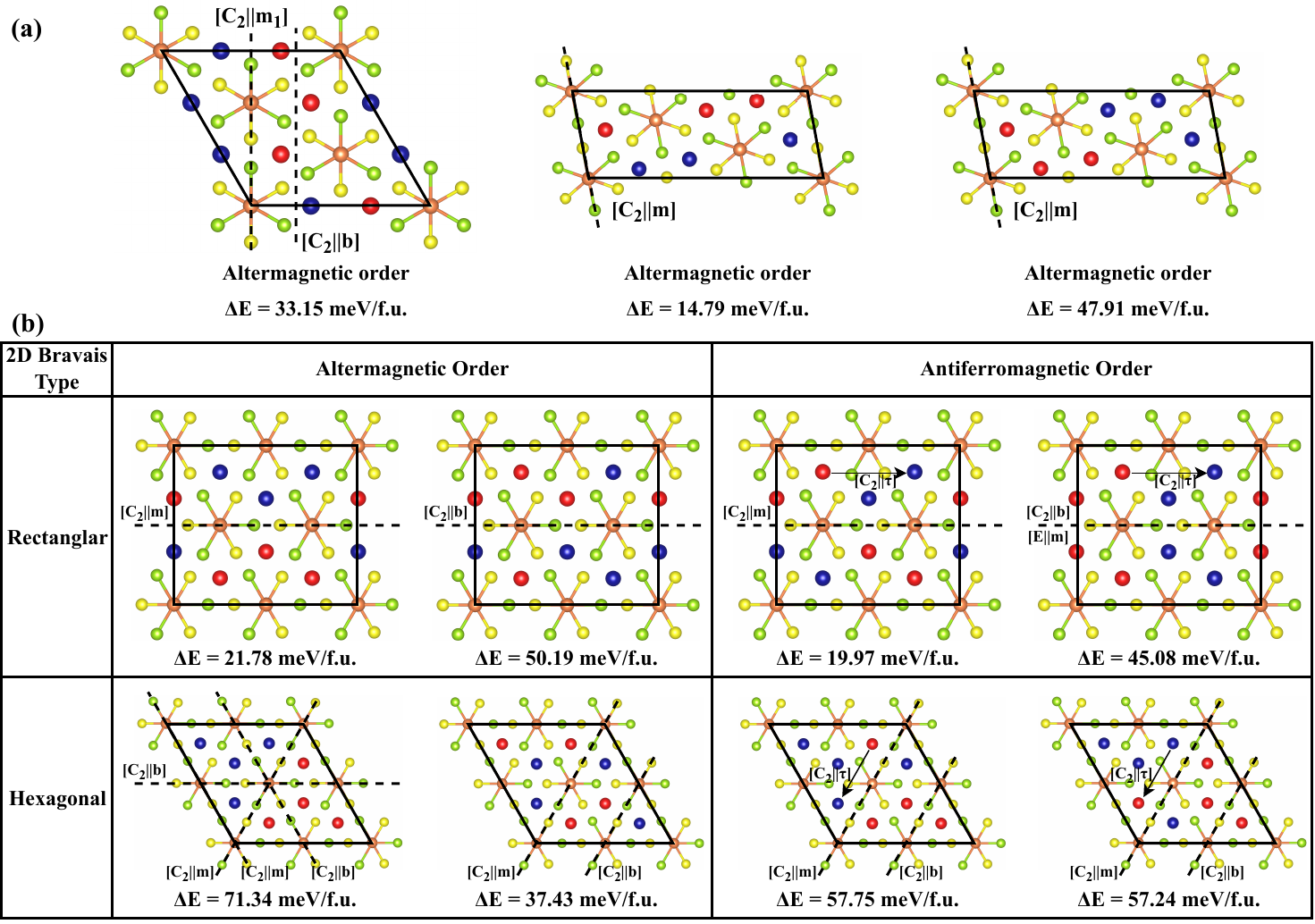}
\caption{Utilizing the methods introduced in Sec. \ref{sec.3}, a greater number of collinear magnetic configurations can be obtained by larger supercells that are three (a) and four (b) times the size of the primitive cell of $\mathrm{Cr}_2 \mathrm{Si}_2 \mathrm{S}_3 \mathrm{Se}_3$. By selecting supercells of different 2D Bravais types, a wider variety of collinear magnetic configurations can be generated. Cyan, blue, and red balls represent Cr atoms, with red and blue balls representing opposite spin sublattices, respectively. Yellow balls represent S atoms, green balls represent Se atoms, and light purple balls represent P atoms. $\Delta E$ denotes the excess energy per formula unit compared to the magnetic ground state. The notation /f.u. represents averaged to each formula unit.}
\label{Fig.5}
\end{figure*}

Utilizing the method introduced in Sec. \ref{sec.3} for generating collinear magnetic configurations, we generated collinear magnetic configurations for the candidate material $\mathrm{Cr}_2 \mathrm{Si}_2 \mathrm{S}_3 \mathrm{Se}_3$ by employing cells that are 2-times and 4-times larger, respectively. The candidate $\mathrm{Cr}_2 \mathrm{Si}_2 \mathrm{S}_3 \mathrm{Se}_3$ (as shown in Fig. \ref{Fig.4}) has the symmetry of group $G_1$, corresponding to $p31m$ (layer group No.70). The group $G_1$ is generated by the set of generators $\langle \tau (1,0,0),\tau (0,1,0),C_{3z},m^{(1)} \rangle$ and can be decomposed into cosets of half group $H_1$ ($p3$, layer group No.65), which is generated by the set of generators $\langle \tau (1,0,0),\tau (0,1,0),C_{3z} \rangle$. $G_1$ orbit of position $x$ occupied by Cr atom can be decomposed into $\{hx \mid h \in H_1 \}$ and $\{gx \mid g \in m^{(1)} H_1 \}$. When up spins occupy $\{hx \mid h \in H_1 \}$ and down spins occupy $\{gx \mid g \in m^{(1)} H_1 \}$, or vice versa, both cases result in $O_{\mid \downarrow \rangle} =m^{(1)} O_{\mid \uparrow \rangle}$ . By iterating through the orbital occupation patterns, we obtain one magnetic configuration denoted as $\mathrm{Cr}_2 \mathrm{Si}_2 \mathrm{S}_3 \mathrm{Se}_3\textendash \mathrm{I}$, shown as Fig. \ref{Fig.4}. 

\begin{figure*}[p]
\centering
\includegraphics[width=\textwidth]{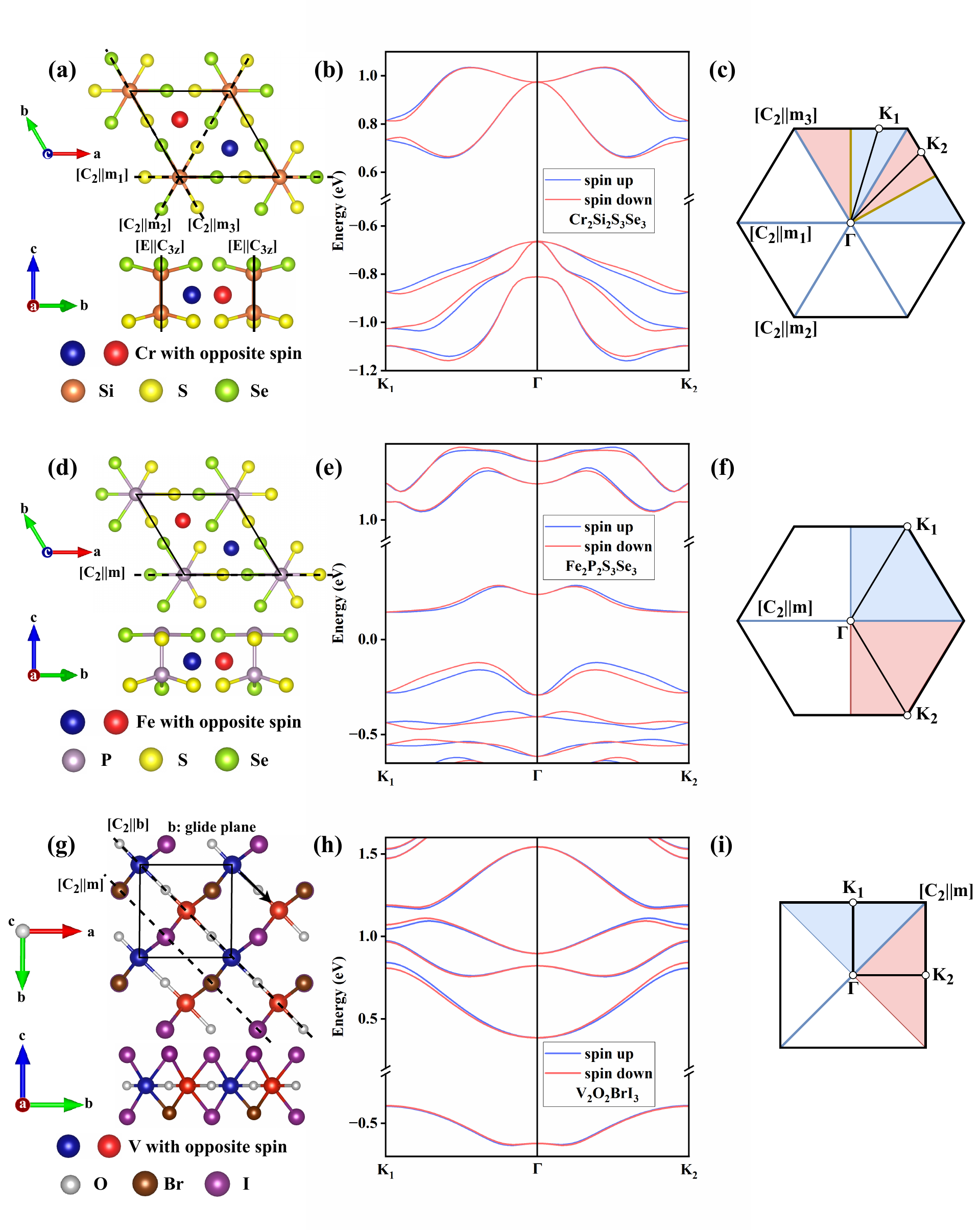}
\caption{The crystal structure and nonrelativistic band structure band structure of $\mathrm{Cr}_2 \mathrm{Si}_2 \mathrm{S}_3 \mathrm{Se}_3$, $\mathrm{Fe}_2 \mathrm{P}_2 \mathrm{S}_3 \mathrm{Se}_3$, and $\mathrm{V}_2 \mathrm{O}_2 \mathrm{Br} \mathrm{I}_3$. The crystal structure of $\mathrm{Cr}_2 \mathrm{Si}_2 \mathrm{S}_3 \mathrm{Se}_3$(a), $\mathrm{Fe}_2 \mathrm{P}_2 \mathrm{S}_3 \mathrm{Se}_3$(d), and $\mathrm{V}_2 \mathrm{O}_2 \mathrm{Br} \mathrm{I}_3$(g), where different colors of magnetic atoms represent the opposite spin sublattices. The band structure of $\mathrm{Cr}_2 \mathrm{Si}_2 \mathrm{S}_3 \mathrm{Se}_3$(b), $\mathrm{Fe}_2 \mathrm{P}_2 \mathrm{S}_3 \mathrm{Se}_3$(e), and $\mathrm{V}_2 \mathrm{O}_2 \mathrm{Br} \mathrm{I}_3$(h) without SOC along the \textbf{k} path $\mathrm{K}_1 \! - \! \Gamma \! - \! \mathrm{K}_2$ is spin splitting, where blue and red solid line represent the opposite spin channels. The \textbf{k} path we use to calculate the band structure of $\mathrm{Cr}_2 \mathrm{Si}_2 \mathrm{S}_3 \mathrm{Se}_3$(c), $\mathrm{Fe}_2 \mathrm{P}_2 \mathrm{S}_3 \mathrm{Se}_3$(f), and $\mathrm{V}_2 \mathrm{O}_2 \mathrm{Br} \mathrm{I}_3$(i). The different colors represent opposite spins.}
\label{Fig.6}
\end{figure*}

\begin{figure*}[t]
\centering
\includegraphics[width=\textwidth]{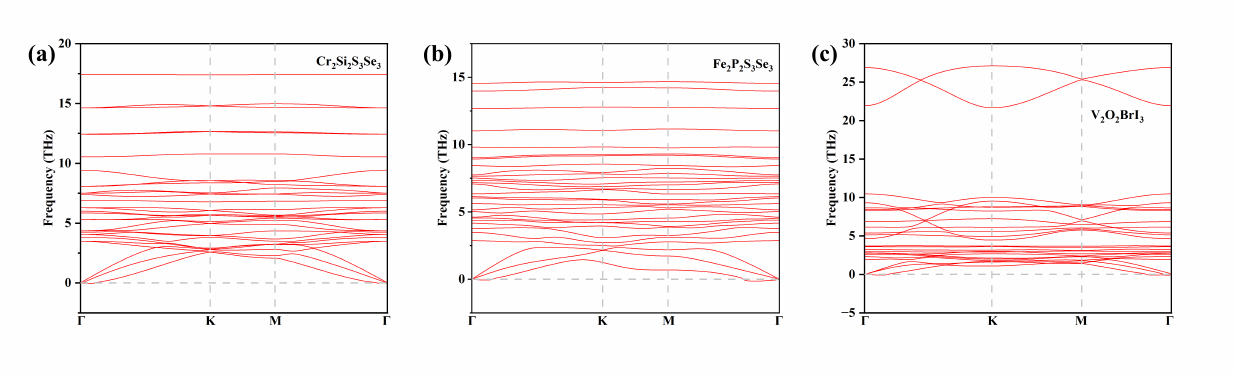}
\caption{Phonon-dispersion spectrum for $\mathrm{Cr}_2 \mathrm{Si}_2 \mathrm{S}_3 \mathrm{Se}_3$(a), $\mathrm{Fe}_2 \mathrm{P}_2 \mathrm{S}_3 \mathrm{Se}_3$(b), and $\mathrm{V}_2 \mathrm{O}_2 \mathrm{Br} \mathrm{I}_3$(c). }
\label{Fig.7}
\end{figure*}

Group $G_1$ has a $t$-subgroup $G_2$, corresponding to $cm11$ (layer group No.13), which is generated by the set of generators $\langle \tau (1,0,0),\tau (1,0,0),\tau (\frac{1}{2},\frac{1}{2},0),m\rangle$, where $m$ is the mirror plane in the $yz$ plane. Subgroup $H_2$ (which correponds to $pb11$, layer group No.12) is generated by the set of generators $\langle \tau (1,0,0),\tau (1,0,0),b\rangle$, where $ b=\tau (\frac{1}{2},\frac{1}{2},0)m$. Due to the set of generators $\langle \tau (1,0,0),\tau (1,0,0),\tau (\frac{1}{2},\frac{1}{2},0),m\rangle$ of $G_2$ can be expressed as $\langle \tau (1,0,0),\tau (1,0,0),m, b\rangle$, $G_2$ can be decomposed into cosets of the half group $H_2$. Employing 2-times cell, $G_2$ orbit of position $x$ occupied by Cr atom can be decomposed into $\{hx \mid h \in H_2 \}$ and $\{gx \mid g \in mH_2 \}$. When up spins occupy $\{hx \mid h \in H_2 \}$ and down spins occupy $\{gx \mid g \in mH_2 \}$, or vice versa, both cases result in $O_{\mid \downarrow \rangle} =mO_{\mid \uparrow \rangle}$ . By iterating through the orbital occupation patterns, we obtain another magnetic configuration denoted as $\mathrm{Cr}_2 \mathrm{Si}_2 \mathrm{S}_3 \mathrm{Se}_3\textendash \mathrm{II}$, as shown in Fig. \ref{Fig.4}. 

In addition, subgroup $H_3$ (which correponds to $pm11$, layer group No.11) is generated by the set of generator $\langle \tau (1,0,0),\tau (1,0,0),m \rangle$. Therefore, $G_2$ can be decomposed into cosets of the half group $H_3$. Employing 2-times cell, $G_2$ orbit of position $x$ occupied by Cr atom can be decomposed into $\{hx \mid h \in H_3 \}$ and $\{gx \mid g \in \tau (\frac{1}{2},\frac{1}{2},0)H_3 \}$. When up spins occupy $\{hx \mid h \in H_3 \}$ and down spins occupy $\{gx \mid g \in \tau (\frac{1}{2},\frac{1}{2},0)H_3 \}$, or vice versa, both cases result in $O_{\mid \downarrow \rangle} =\tau (\frac{1}{2},\frac{1}{2},0)O_{\mid \uparrow \rangle}$ . By iterating through the orbital occupation patterns, we obtain another magnetic configuration denoted as $\mathrm{Cr}_2 \mathrm{Si}_2 \mathrm{S}_3 \mathrm{Se}_3\textendash \mathrm{III}$, shown as Fig. \ref{Fig.4}. 

Based on the analysis of the candidates $\mathrm{Cr}_2 \mathrm{Si}_2 \mathrm{S}_3 \mathrm{Se}_3$, the sublattice $O_{\mid \uparrow \rangle}$ occupied by up spins in the magnetic configuration $\mathrm{Cr}_2 \mathrm{Si}_2 \mathrm{S}_3 \mathrm{Se}_3\textendash \mathrm{I}$ cannot obtain the opposite sublattice $O_{\mid \downarrow \rangle}$ through inversion operation $\bar{E}$, translation operation $\tau$, rotation operation $C_{2z}$ around an axis parallel to the $z$ axis, mirror operation $m_z$ parallel to the $xy$ plane, or their products with translation operation $\tau$. Therefore, the magnetic configuration $\mathrm{Cr}_2 \mathrm{Si}_2 \mathrm{S}_3 \mathrm{Se}_3\textendash \mathrm{I}$ possesses altermagnetism. In the cases of $\mathrm{Cr}_2 \mathrm{Si}_2 \mathrm{S}_3 \mathrm{Se}_3\textendash \mathrm{II}$ and $\mathrm{Cr}_2 \mathrm{Si}_2 \mathrm{S}_3 \mathrm{Se}_3\textendash \mathrm{III}$, the sublattice $O_{\mid \uparrow \rangle}$ occupied by up spins can obtain the opposite sublattice $O_{\mid \downarrow \rangle}$ through performing a translation symmetry operation $\tau ( \frac{1}{2},\frac{1}{2},0)$. Hence, the magnetic configurations $\mathrm{Cr}_2 \mathrm{Si}_2 \mathrm{S}_3 \mathrm{Se}_3\textendash \mathrm{II}$ and $\mathrm{Cr}_2 \mathrm{Si}_2 \mathrm{S}_3 \mathrm{Se}_3\textendash \mathrm{III}$ possesses antiferromagnetism. 

Based on the 3-times and 4-times larger supercell, we obtain a greater variety of collinear magnetic configurations. In particular, by choosing supercells of different 2D Bravais types, we can produce a more diverse range of collinear magnetic configurations. Through the aforementioned methods, we generated a series of collinear magnetic configurations with opposite sublattices connected by symmetry operations. After eliminating magnetic configurations that are equivalent, we are left with 12 unique configurations, as shown in Fig. \ref{Fig.5}. First-principles calculations confirm that the altermagnetic configuration $\mathrm{Cr}_2 \mathrm{Si}_2 \mathrm{S}_3 \mathrm{Se}_3\textendash \mathrm{I}$ is the magnetic ground state, with an energy per formula unit lower than other magnetic configurations, and 90.32 meV lower compared to the ferromagnetic configuration. Therefore, we obtain the structure $\mathrm{Cr}_2 \mathrm{Si}_2 \mathrm{S}_3 \mathrm{Se}_3$ with two-dimensional altermagnetism. The crystal structure and nonrelativistic band structure of $\mathrm{Cr}_2 \mathrm{Si}_2 \mathrm{S}_3 \mathrm{Se}_3$ are shown in Fig. \ref{Fig.6}(a). Utilizing the similiar approach, we generate two-dimensional material $\mathrm{Fe}_2 \mathrm{P}_2 \mathrm{S}_3 \mathrm{Se}_3$ with altermagnetism, as shown in Fig. \ref{Fig.6}(d). 

In addition to $\mathrm{Cr}_2 \mathrm{Si}_2 \mathrm{S}_3 \mathrm{Se}_3$ and $\mathrm{Fe}_2 \mathrm{P}_2 \mathrm{S}_3 \mathrm{Se}_3$ of hexagonal lattice, we generate two-dimensional material $\mathrm{V}_2 \mathrm{O}_2 \mathrm{Br}_{4x} \mathrm{I}_{4(1-x)}$ with altermagnetism of rectangular lattice based on the same method. The symmetry of two-dimensional magnetic materials $\mathrm{V} \mathrm{O} \mathrm{Br}_2$ and $\mathrm{V} \mathrm{O} \mathrm{I}_2$ corresponds to layer group $pm2m$ (No.27). The $pm2m$ layer group has subgroups $p11m$, $p211$, and $pm11$. Corresponding to different subgroups $H_k$, the positions of Br and I atoms in a $\sqrt{2} \times \sqrt{2} \times 1$ supercell are divided into different $H_k$ orbits. Allow components to occupy all atomic positions on the orbits and generate candidate structures with specific symmetries $H_k$. After removing equivalent structures, we obtain 7 candidates. We generate possible magnetic configurations of candidates $\mathrm{V}_2 \mathrm{O}_2 \mathrm{Br}_{4x} \mathrm{I}_{4(1-x)}$. In order to determine whether these structures are possible to possess altermagnetism, we analyze the symmetry operations that connect the opposite sublattices. Then, candidates $\mathrm{V}_2 \mathrm{O}_2 \mathrm{Br} \mathrm{I}_3$, $\mathrm{V}_2 \mathrm{O}_2 \mathrm{Br}_2 \mathrm{I}_2$, and $\mathrm{V}_2 \mathrm{O}_2 \mathrm{Br}_3 \mathrm{I}$, which may possess altermagnetism, are retained for first-principles calculations. Through first-principles calculations, it was determined that the magnetic ground states of the candidates $\mathrm{V}_2 \mathrm{O}_2 \mathrm{Br} \mathrm{I}_3$ (shown in Fig. \ref{Fig.6}(g)), $\mathrm{V}_2 \mathrm{O}_2 \mathrm{Br}_2 \mathrm{I}_2$, and $\mathrm{V}_2 \mathrm{O}_2 \mathrm{Br}_3 \mathrm{I}$ are altermagnetic. The magnetic ground state of the candidate $\mathrm{V}_2 \mathrm{O}_2 \mathrm{Br} \mathrm{I}_3$ with an energy per formula unit 13.24 meV lower than that of the ferromagnetic order, the magnetic ground state of the candidate $\mathrm{V}_2 \mathrm{O}_2 \mathrm{Br}_2 \mathrm{I}_2$ with an energy per formula unit 28.67 meV lower, and the magnetic ground state of the candidate $\mathrm{V}_2 \mathrm{O}_2 \mathrm{Br}_3 \mathrm{I}$ with an energy per formula unit 31.04 meV lower. 

To evaluate the dynamic stability, we calculated the phonon spectrum of the two-dimensional altermagnetic structures $\mathrm{Cr}_2 \mathrm{Si}_2 \mathrm{S}_3 \mathrm{Se}_3$, $\mathrm{Fe}_2 \mathrm{P}_2 \mathrm{S}_3 \mathrm{Se}_3$, and $\mathrm{V}_2 \mathrm{O}_2 \mathrm{Br} \mathrm{I}_3$, as shown in Fig. \ref{Fig.7}. The absence of any negative phonon frequencies confirms the dynamical stability of the structures $\mathrm{Cr}_2 \mathrm{Si}_2 \mathrm{S}_3 \mathrm{Se}_3$, $\mathrm{Fe}_2 \mathrm{P}_2 \mathrm{S}_3 \mathrm{Se}_3$, and $\mathrm{V}_2 \mathrm{O}_2 \mathrm{Br} \mathrm{I}_3$ in our predictions. 

\section{SUMMARY}

In summary, we have provided a general approach to generating multi-component structures with two-dimensional altermagnetism,  as well as a systematic method for generating collinear magnetic configurations where opposite-spin lattices are interconneccted through symmetry operations. While this study primarily focuses on the application of generating candidate structures with specific symmetries and collinear magnetic configurations in two-dimensional altermagnets, the proposed methodology is equally applicable to three-dimensional systems. We anticipate that this approach will significantly contribute to the discovery of novel altermagnetic materials in future research. Meanwhile, we identified several stable two-dimensional altermagnetic materials, validating the use of this approach in the search for 2D altermagnets. Furthermore, this approach generates a rich variety of collinear magnetic configurations that meet symmetry requirements, improving the accuracy of magnetic ground state predictions. Among the series of altermagnetic materials, we have provided structures with dynamic stability, providing a reference for the experimental synthesis of two-dimensional altermagnetic materials. 

\begin{acknowledgments}
This work is supported by the Guangdong Basic and Applied Basic Research Foundation (Grants No. 2023A1515110894), the National Natural Science Foundation of China (Grant No. 12074126, No. 12474228). This work is partially supported by High Performance Computing Platform of South China University of Technology. 
\end{acknowledgments}

\appendix

\section{COMPUTATIONAL DETAILS}

All calculations were performed using the Vienna Ab initio Simulation Package (VASP) \cite{KRESSE199615,PhysRevB.54.11169}, employing the projector augmented wave method \cite{PhysRevB.59.1758} based on density functional theory. A cutoff energy of 500 eV was set for the plane wave basis. The structure was relaxed until the forces on atoms were below 0.001 eV/$\mathrm{\AA}$ and the convergence criterion was $1\times10^{-7}$ eV for the energy difference in the electronic self-consistent calculation. A vacuum of 12$\mathrm{\AA}$ was constructed perpendicular to the material plane. The SOC effect was not considered in the calculations.

\nocite{*}


%

\end{document}